\documentclass[twocolumn,showpacs,preprintnumbers,amsmath,amssymb,aps]{revtex4}

\usepackage{bm}%
\usepackage{graphicx}

\begin{document}

\title{The window at the edge of chaos in a simple model of gene interaction networks}

\author{Dejan Stoki\'c$^1$}
%\email{dejan.stokic@meduniwien.ac.at}
\author{Rudolf Hanel$^1$}
%\email{rudolf.hanel@meduniwien.ac.at}
\author{Stefan Thurner$^{1,2}$}
\email{thurner@univie.ac.at}
\affiliation{$^{1}$Complex Systems Research Group, HNO, Medical University of 
Vienna, W\"ahringer G\"urtel 18-20, A-1090 Vienna\\
 $^{2}$Santa Fe Institute, 1399 Hyde Park Road, Santa Fe, NM 87501, USA }

\begin{abstract}
As a model for gene and protein interactions we study a set for molecular catalytic reactions. 
The model is based on experimentally motivated interaction network topologies, and is designed to capture  
some key  statistics of gene expression statistics. 
We impose a non-linearity to the system by a boundary condition which guarantees non-negative 
concentrations of chemical concentrations and  study the system stability quantified by maximum Lyapunov exponents. 
We find that the non-negativity constraint leads to a drastic inflation of those regions in parameter space where the  
Lyapunov exponent exactly  vanishes. We explain the finding as a self-organized critical phenomenon. 
The robustness of this finding with respect to different network topologies and  the role of intrinsic 
molecular- and external noise is discussed. 
We argue that systems with inflated 'edges of chaos' could be much more easily favored by 
natural selection than systems where the Lyapunov exponent vanishes only on a parameter set of measure zero. 
\end{abstract}
%\date{\today}

\maketitle

\section{Introduction}

Most complex systems, living systems in particular,  are characterized by remarkable degrees of 
stability and at the same time by a tremendous potential of flexibility and adaptability. This has led some authors to define 
complex and living systems as living at the 'edge of chaos', \cite{langton,kauffman_order,mitchell,packard} meaning 
--in a somewhat picturesque way-- that it takes only tiny 
changes in the system to move it from a stable and regular mode into a chaotic phase where large portions of 
phase space can get sampled. 
The concept is that systems at the edge of chaos are especially well suited for adaptation and 
information processing in the sense that adaptability is associated by the possibility of finding adequate 
new states in possibly changed environments at very fast rates.
It has been argued that living systems at the edge of chaos would get favored by natural selection, and that 
life has evolved towards such a special region in parameter space \cite{kauffman_order}.
In many dynamical systems the edge of chaos is a very special set of points in parameter space,
often of measure zero, 
characterized by the system's maximal Lyapunov exponent $\lambda$ passing through zero.
It is not clear how systems can get regulated towards (or have evolved towards) 
such a limited set of critical points, even though 
some interesting ideas have been proposed in this direction \cite{melby}. 
Even in the simplest maps like the logistic map, the dynamics exactly at these special points  can 
become highly non-trivial \cite{robledo}.   
 
It is evident that living systems have evolved towards stable systems in stationary disequilibrium. 
Various authors argue that a key principle of living systems is their ability to replicate \cite{eigen};  
corresponding rate equations for molecular replicators have been proposed for a  long time, beginning with \cite{lotka}. 
As such, basic molecular reactions in living systems (e.g. protein production or degradation) have to be autocatalytic. 
If autocatalytic reactions are not balanced by degradation and/or thermostatic net-flow of substance 
to and fro the system (like in a flow reactor), concentrations of molecular products  will diverge in the replicator.
A stationary state can be established when production and decay (flow) rates of inter cellular molecules effectively 
balance each other,  \cite{pross,pross2}. In this sense stability (stationarity) provides a natural selection criterion. 
Catalytic reactions are simply described by reaction networks, which contain production and degradation 
rates. Given current developments in genomics- and proteomics technology some facts about these networks 
become known. 
By now there is some evidence that these (directed) networks show scale-free  (SF) topological organization 
\cite{sneppen,jeong2}.  On the basis of a given  molecular reaction topology \cite{jeong}
several  gene network models have been proposed \cite{wolf,holte,kikuchi}. In principle  
two different approaches have been pursued: discrete approaches, using Boolean networks \cite{kauffman} and continuous approaches, using ordinary or stochastic differential equations \cite{smith,mahaffy,mestl,chen}. 
Combinations of both have also been reported \cite{mcadams,yoshida}. 

Recently the importance  of noise in molecular reaction networks has been stressed and 
its relevance has been experimentally demonstrated \cite{ko,fiering,hasty}. For example the level of noise can 
determine whether cells in Drosophila become epidermal or neural cells \cite{haitzler}. Further it was shown that 
low reproduction rates of DNA and important regulatory molecules forbid to 
neglect stochastic effects \cite{guptasarma}. Intrinsic noise, microscopic events within the cell, 
and extrinsic noise, such as cell to cell variations, are experimentally distinguishable \cite{elowitz}.
In this context  a stochastic differential equation model  has been proposed for 
regulatory transcription networks \cite{chen2}. 

In this work we study a simple linear, noise driven dissipative model for 
catalytic molecular reactions. We impose a non-linearity to this system by assuming that concentrations 
can not become negative. We demonstrate that this non-linearity changes the 'the edge of chaos'
from a point where $\lambda=0$, to extended regions  of vanishing Lyapunov exponents. 
The model offers a way to understand how systems naturally evolve and adapt towards a widened 'edge of chaos'.

\section{The Model}

We assume that gene to gene interactions can be modeled as chemical reactions between 
proteins, mRNA and other nucleic material.   
Let us denote the concentrations of proteins $i$ at time $t$ by the vector $p_i(t)$ and of RNA molecules $j$ by $r_j(t)$.
For convenience let us combine all types of concentrations into a single vector $x\equiv (\vec p, \vec r)$. The 
size of the vector (number of products) we denote by $N$.
The simplest linear model to capture all possible interactions is given by 
\begin{equation}
\frac{d}{dt} x_i = \sum_j A^0_{ij} x_j \quad , 
\end{equation}
where $A^0$ is the matrix of reaction rates. Even though this model is clearly an over-simplification   
of reality it has been frequently used recently \cite{yeung,holte,dhaeseler}. 
Let us assume that these rates are not perfect constants but fluctuate according to $A^0_{ij}(t)=A_{ij}+\xi_{ij}(t)$, 
for example through thermal noise. For simplicity let $\xi_{ij}$ be an iid process with zero mean.
Replacing $A^0$ by $A$ we get 
\begin{equation}
\frac{d}{dt} x_i = \sum_j A_{ij} x_j  + \xi_{ii}(t) x_i +\sum_{j\neq i}  \xi_{ij} (t) x_j \quad .  
\label{model_raw}
\end{equation}
Regardless of the distribution of $\xi_{ij} $, and assuming that $x$ will converge to a 
reasonably stationary distribution, according to the central limit theorem, 
the sum of the right hand side will yield a random number from a Gaussian distribution, which
we shall denote by $\eta_i \in N(0,\sigma)$. For simplicity we shall assume $\xi_{ii}$ also Gaussian, i.e. 
$\xi_{ii} \equiv \xi_{i}   \in N(0,\bar \sigma)$with the same variance $\bar \sigma $ $\forall i$.  
We need a final addition of our model Eq. (\ref{model_raw}), to incorporate a further 
experimental observation. Many gene-products (e.g. mRNA levels) fluctuate 
around some characteristic value, $x^0_i$, across the cell cycle.
In Fig. \ref{tsallisfig}a we show the expression levels for mRNA levels of 10 randomly picked 
genes from the yeast genome (\textit{S.cerevisae}) over 2 cell cycles at 17 time points 
taken at 10 minute intervals \cite{cho, zivkovic}. The number of measured genes in 
the budding yeast genome was $N=6220$. 
To incorporate these characteristic 
levels we chose values $x^0_i$ from some distribution. In Fig \ref{tsallisfig}b we show the 
experimental distribution of mRNA expression levels of the yeast genome, defined as the time-average 
over cell cycles. 
$x^0_i=\langle x_i(t) \rangle_t$. In the following  $x^0_i$ is taken from uniform, Gaussian or 
the experimental distribution. This fixes our model to be
\begin{equation}
\frac{d}{dt} x_i = \sum_j A^0_{ij}( x_j - x^0_j) + \xi_{i}(t) x_i + \eta_i \quad ,   
\label{model}
\end{equation}
with $\xi_i\in N(0,\bar \sigma)$ and $\eta_i\in N(0, \sigma)$ 
the multiplicative and additive noise components respectively.
Multiplicative and additive noise can be interpreted as 
intrinsic and extrinsic noise as used e.g. in \cite{elowitz}. Both intrinsic and extrinsic noise 
are present in gene networks.

To be able to interpret $x$ as concentrations we have to introduce the constraint 
\begin{equation}
x_{i}(t) \geq 0  \quad \forall i \quad,
\label{poshalfsp}
\end{equation}
which means that regardless of $\dot x_i$  of (\ref{model}), $x_i(t)$ can never be below 
zero. This imposes a non-linearity onto the  system  and makes it non-trivial.

\subsection{The interaction matrix}

Before solving the system we have to specify the interaction 
network, i.e. the matrix elements (chemical rates) of $A$. It is obvious that the network will 
be directed and weighted. Diagonal elements $A_{ii}<0$ are decay rates of the products, 
off-diagonal rates $A_{ij}$ can be both positive or negative, corresponding to activation or inhibition.
Further, it is clear that first  not all products can interact with each other, i.e. 
a large number of matrix elements will be zero, and second most rates are not available 
from experiments. We are thus led to model $A$ as a random matrix in the following way. 
Using terminology from network theory the 'degree' $k_i$ of product $i$ is defined as 
the number of products that can be regulated by product $i$. 
The class of interaction networks can now be  specified by the 'degree distribution'. 
There is evidence that protein networks \cite{jeong2} and metabolic networks  \cite{jeong}
are scale-free networks $p(k)\sim k^{-\gamma }$, characterized by a degree distribution with average
degree  $\langle k\rangle > 4$, and an exponent $\gamma \sim 2.2$. In the following we generate 
such networks and contrast them to classical random networks, i.e. Erd\"os-Renyi graphs (ER) \cite{er1959,er1960}
with the same average degree. If the number of non-zero rates in $A$ is denoted by $L$, the 
average connectivity is $\langle k\rangle= L/N$.

Once it is decided which products interact with each other, i.e.  $A_{ij}\neq0$, the actual rates have to be fixed. 
We assume these being Gaussian, $A_{ij}\in N(0,\sigma_A)$. This  is supported  experimentally e.g. by \cite{dhaeseler},
where a least squares fit of synthetic gene network models to real data indicates 
that the normal distribution of interaction weights gives the best results.
The (negative) decay rates $A_{ii}$ we take constant and equal  $\forall i$ in  the following.

\subsection{A note on multiplicative noise}

Note that the diagonal component of Eq. (\ref{model}) -- ignoring the positivity constraint -- immediately  reminds at the 
stochastic differential equation, 
\begin{equation}
\frac{d}{dt} x = f(x) + g(x)\xi(t) + \eta(t) \quad ,
\label{multlangevin}
\end{equation}
with $f=-A_{ii}x$ and $g(x)=x$. This Langevin process has been exactly solved 
\cite{anteneodo}, the solution (probability distribution of $x$) being a $q$-exponential, 
\begin{equation}
p(x)\sim \left [ 1+ (q-1) \beta x^2    \right]^{1/(1-q)}  \quad , 
\label{tsallis}
\end{equation} 
with $\beta=(-A_{ii}+\bar \sigma /2 )/\sigma$, 
where $(1-q)^{-1}$ is the asymptotic power exponent. 
In Fig. 1c we show experimental data confirming the power law aspect of 
mRNA concentrations in yeast in the same data \cite{cho}. 
$\Delta x_i(t)\equiv x_i(t)- x_i(t-1)$ is the difference in gene expression levels between two consecutive measurements; 
$P(>\Delta x)$ is the cumulative distribution, for all $i$ and $t$.
In the same plot we show results of a numerical simulation of the model Eq. (\ref{model}) with the 
$N=6000$ ER topology for $A$ at $\langle k \rangle=20$, for the two cases:  first, $\bar \sigma=0$ and   $ \sigma>0$ (Gaussian noise model), 
and second $\bar \sigma>0$ and   $ \sigma=0$ (multiplicative noise model). 
Further a $q$-exponential fit \footnote{The cumulative pdf of Eq. (\ref{tsallis}) is again a $q$-exponential, 
with a different value, $q_{\rm cum}$.}
to the data is shown (broken line), with an effective $q_{\rm cum}=1.55$ . 
\begin{figure}
\begin{tabular}{c}
\hspace{-0cm} \resizebox{0.40\textwidth}{!}{\includegraphics{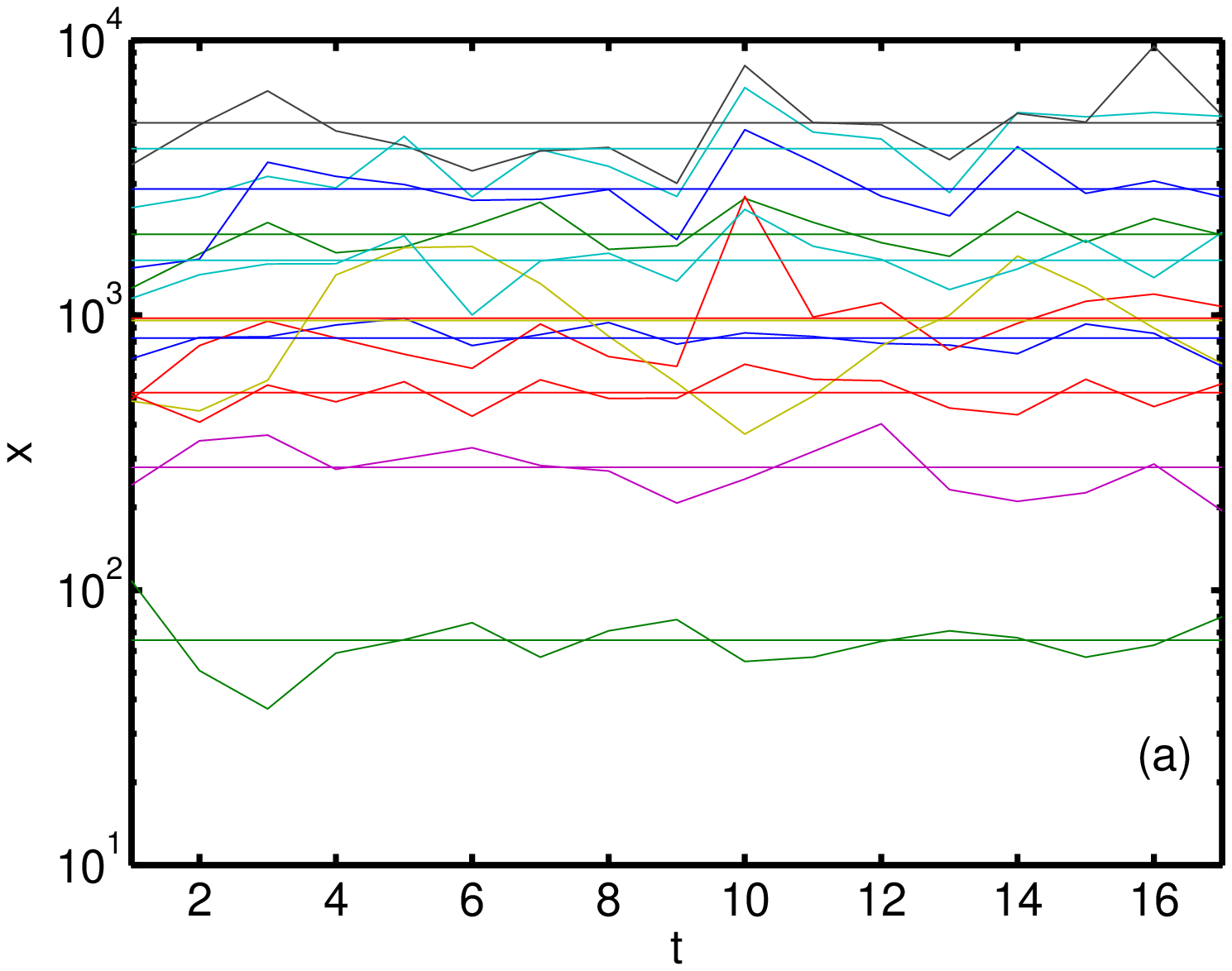}}\\
\hspace{-0cm} \resizebox{0.40\textwidth}{!}{\includegraphics{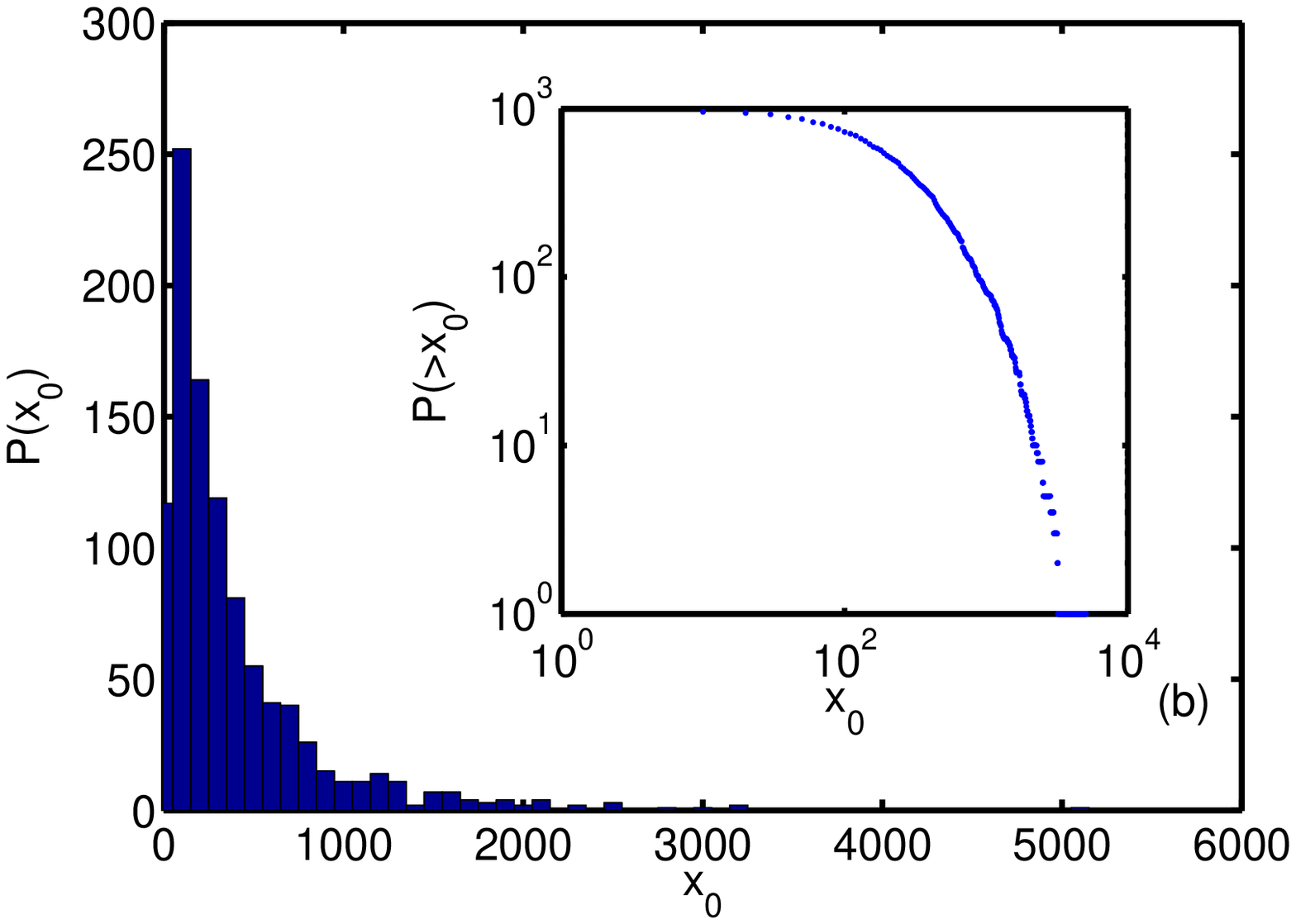}}\\
\hspace{-0cm} \resizebox{0.40\textwidth}{!}{\includegraphics{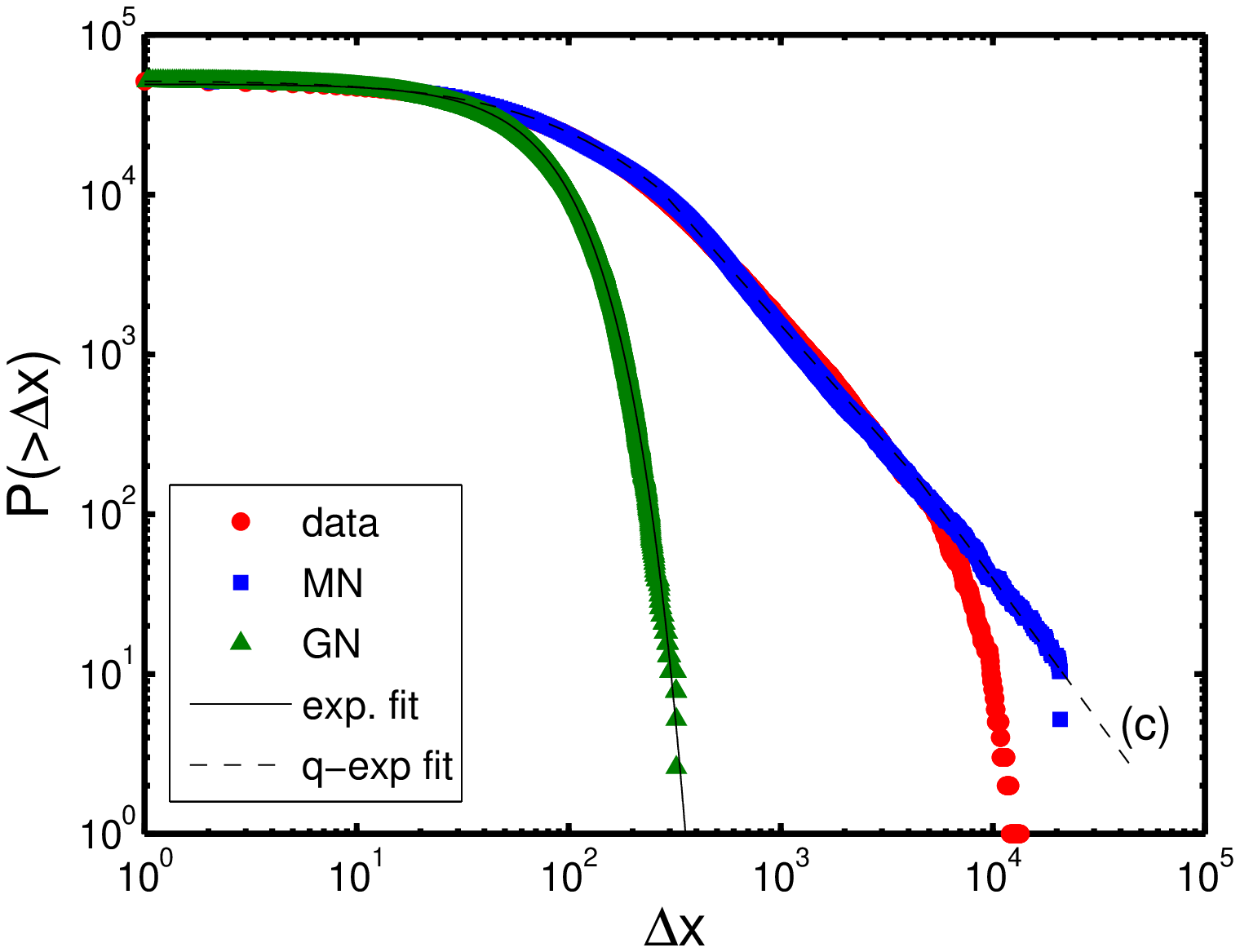}}\\
\end{tabular}
\caption{
(a) A set of 10 randomly picked gene-expression trajectories of yeast (\textit{S. cerevisiae}) over two cell cycles \cite{cho}. 
 Stationary state values $x^{0}$   are defined as the time-averages of gene expression 
levels. $x_0 $ levels corresponding to the shown genes are indicated by horizontal lines.  
(b) Distribution of stationary state values $x^{0}$. Inset shows the cumulative distribution.
(c) Cumulative distribution of gene expressions increments for the same data (circles) 
for the numerical simulation of the evolution of gene expression data with two 
evolution models with additive Gaussian (boxes) and multiplicative (diamonds) external noise. }
\label{tsallisfig}
\end{figure}

\subsection{Stability}

Biological systems are sufficiently stable, i.e. products are not produced ad infinitum, and at the same time 
are sufficiently dynamical. If our model is reasonable 
we have to find the regions in parameter-space where such non-trivial stability is ensured. 
If we ignore for a moment the positivity condition, Eq. (\ref{poshalfsp}), 
and the stochastic terms in Eq. (\ref{model}), the stability of the system 
is dominated by the largest real part of the eigenvalues of the interaction matrix $A$. 
If there are no non-negative real parts of the eigenvalues, the system will be asymptotically stable. 
If the distribution of off-diagonal elements in $A$ is normal \cite{dhaeseler} with variance $\sigma^2_A$, 
the eigenvalue spectrum is -- according to a powerful result from random matrix theory --
a circle in the complex plane (Girko's circular law), \cite{girko}.
For a fully connected matrix, with $L=N^2$ non-zero entries in $A$ 
the radius of this circle $\rho$  is equal to the product of the 
standard deviation and square root of the system size $N$.
For non-fully connected networks, $L<N^2$, the radius is given by
(see e.g. \cite{crisanti_book})
\begin{equation}
\rho= \sigma_A \sqrt{L/N} = \sigma_A \sqrt{\langle k\rangle}\quad . 
\label{girkorad}
\end{equation} 
If the diagonal elements of the random matrix are from a zero-mean distribution, 
Girko's circle is centered at the origin of the complex plane. 
In our case we have $A_{ii}<0$ and the center of the circle will be shifted to the position 
$(-A_{ii},0)$,  see e.g. \cite{bielyQF07}. 

Now, including the positivity constraint and the stochastic dynamics, 
it is obvious that the eigenvalue spectrum of $A$ will only be a part of the story. 
To define a measure for  system stability that captures these aspects, 
maybe the simplest choice is the maximal Lyapunov exponent 
\begin{equation}
\lambda\equiv\lim_{t \to \infty}\frac{1}{t}\ln\left( \frac{\Vert\delta x(t)\Vert}{\Vert\delta x(0) \Vert}\right)\quad, 
\label{lyapunovexp}
\end{equation}
where $\delta x(t) \equiv x(t)-x'(t)$, is the difference between two trajectories, where 
$x'(t)$ results from a small perturbation in the initial condition, $\Vert\delta x(0)\Vert\ll1$.  

For the system without the positivity condition the $\lambda$ can now be related to $\rho$
in the following way
\begin{equation}
\lambda(\langle k \rangle) \sim  \rho(\langle k \rangle)   -  A_{ii}  = \sigma_A  \sqrt{\langle k\rangle}  - A_{ii}  \quad, 
\label{maxlyap}
\end{equation} 
where the $A_{ii}$ is the eigenvalue spectrum shift discussed above. 

For the case of the full model i.e. with the positivity condition,  we hypothesize the following 
scenario: With strong noise levels after some time  several  trajectories diffuse to 
hit zero. For long enough times the expected number of these trajectories will be 
half of the total number, $N/2$. This amounts to a reduction of system size by one half 
(with fixed connectedness), i.e.
$N \to N_{\rm eff }=N/2$ and $L\to L_{\rm eff }=L/4 $. For connectivity this means, 
$\langle k \rangle\to \langle k_{\rm eff } \rangle= \langle k \rangle/2$. We would thus expect the asymptotic 
(large $\langle k \rangle$) behavior 
of $\lambda$ of the {\em full} model as a function of connectivity
\begin{equation}
\lambda(\langle k \rangle) \sim   \sigma_A  \sqrt{\frac {\langle k\rangle}{2} }  - A_{ii}  \quad, 
\label{maxlyap2}
\end{equation}
For small $\langle k \rangle$ where no (or few) trajectories hit zero, of course, we expect Eq. 
(\ref{maxlyap}) to hold. More generally, it is reasonable to assume that for 
given connectivity and noise levels, there will be an effective connectivity, 
\begin{equation}
\langle k_{\rm eff } \rangle= \langle k \rangle N_{\rm on }/N  \quad ,
\label{maxlyap3}
\end{equation} 
where $N_{\rm on }$ is the number of trajectories not  at zero. 

Let us finish with commenting on a potential stabilizing role of  
multiplicative noise \cite{khas}. Consider the one dimensional case of our model 
\begin{equation}
\frac{d}{dt} x = a(x-x_{0}) +  \xi x + \eta \quad, 
\label{nstocheqspec}
\end{equation}
with $\xi\in N(0,\bar  \sigma)$ and $\eta \in N(0, \sigma)$. 
The evolution of a perturbation $\delta x$ thus follows 
\begin{equation}
\frac{d}{dt} \delta x = a \delta x +  \xi \delta x \quad, 
\label{onedperteq}
\end{equation}
with the solution
\begin{equation}
\delta x(t) = \delta x(0)e^{(a - \frac{\bar \sigma^{2}}{2})t} e^{\bar \sigma \int dt \xi (t) } \quad .
\label{onedpertsol}
\end{equation}
The Lyapunov exponent is proportional to $a - \bar \sigma^{2}/ 2$ 
showing  that the system can be stable even for positive $a$.

\section{Results}

We numerically solve the model Eq. (\ref{model}), now  with positivity condition and compare with 
the above predictions. 
For numerical simulations we generated scale-free (SF) and ER  networks of sizes 
of $N=200,500,1000$. To vary $\langle k \rangle$ we adjusted the number of 
non-zero rates $L$ in the matrix $A$. For scale-free networks we fixed the scaling exponent $\gamma=2.2$. 
All the following results are averages over 20 
random realizations of networks for a given parameter set. 
The Lyapunov exponents were computed from datasets of 1000 timepoints, after 
discarding the initial 200 timesteps. $x_0$ was chosen from 
the experimental distribution of Fig. \ref{tsallisfig}b. We do not  observe noteworthy 
changes of results when using uniform or  Gaussian distributions. 
\begin{figure}
\begin{center}
\hspace{-0cm} \resizebox{0.40\textwidth}{!}{\includegraphics{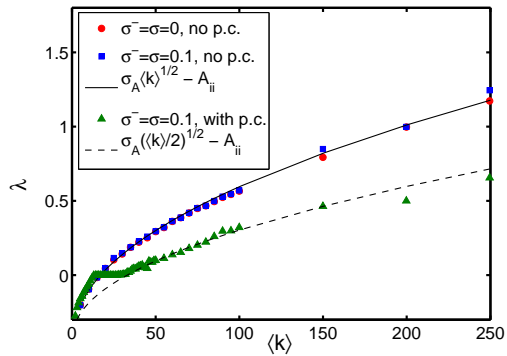}} 
\end{center}
\caption{Maximum Lyapunov exponents $\lambda$ as a function of average degree $\langle k \rangle$, 
averaged over 20 realizations, for  ER networks. The influence of the positivity condition on forming a plateau is 
immediately seen. Simulations are shown with ($\sigma=\bar \sigma=0.1$) and without ($\sigma=\bar \sigma=0$) 
noise, $N=500$, $A_{ii}=-0.4$. Lines are Eqs. (\ref{maxlyap})  and  (\ref{maxlyap2}).}
\label{woPC}
\end{figure}
In Fig.  \ref{woPC} we show the solution for $\lambda$ for the ER network,
as a function of $\langle k \rangle$, with (triangles) and  without (circles) positivity condition. 
The corresponding theoretical predictions Eqs. (\ref{maxlyap}) and (\ref{maxlyap2}), 
are drawn as broken and solid lines, respectively. The case without  positivity 
condition is completely explained by theory over the entire range of $\langle k \rangle$, Eq. (\ref{maxlyap}).  
In the case with the constraint  the asymptote follows Eq. (\ref{maxlyap2}), as expected. 
It is also seen that for small $\langle k \rangle$, Eq. (\ref{maxlyap}) is valid. 
The main finding of this paper is that within a $\langle k \rangle$-window between about 
10 and 30 a plateau forms where $\lambda$ practically vanishes. 
The constraint dynamics allows a full range of connectivities to support a 'life at the edge of chaos'.  

\begin{figure}[t]
\begin{center}
\hspace{-0cm} \resizebox{0.30\textwidth}{!}{\includegraphics{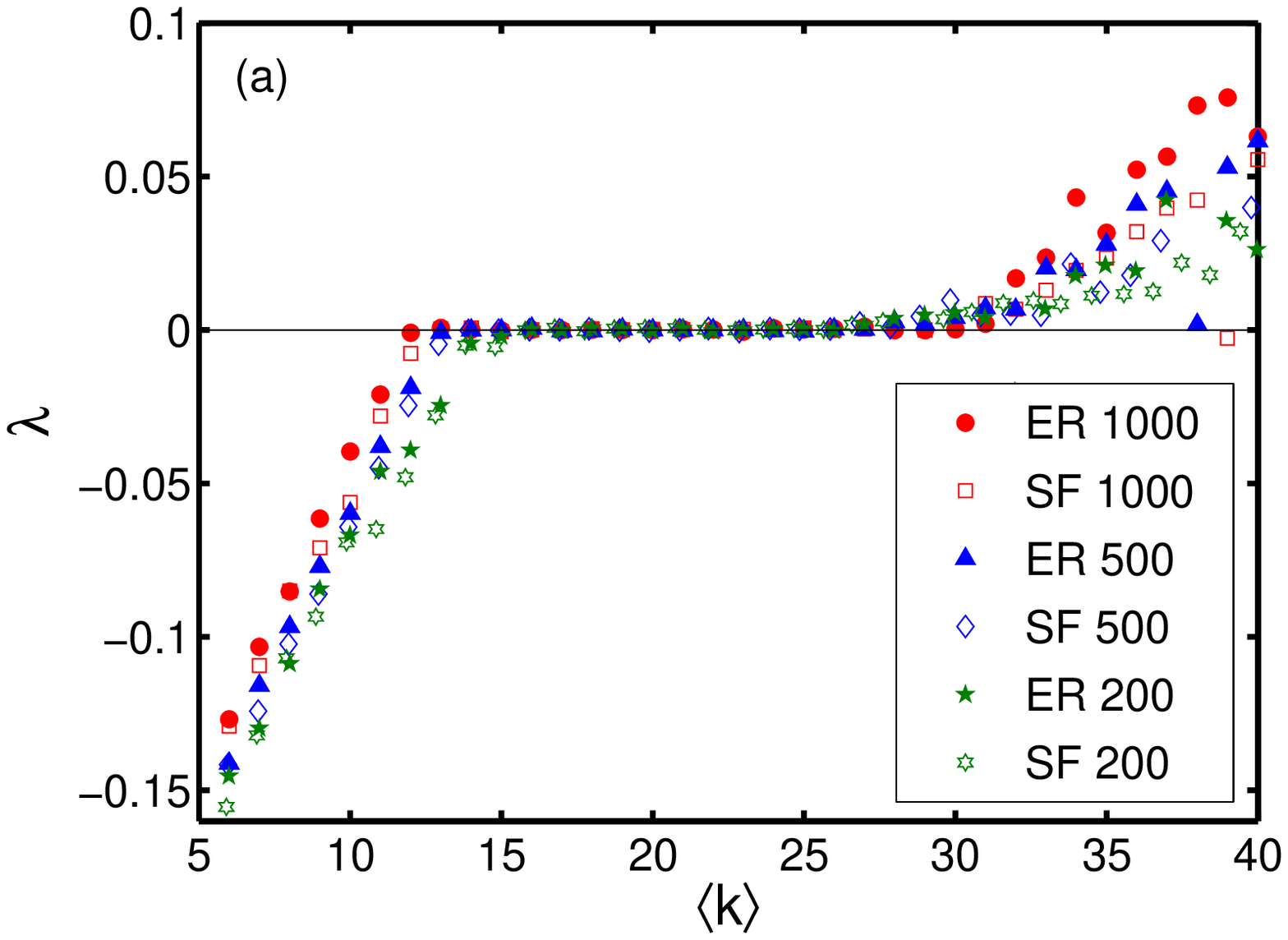}} \\
\hspace{-0cm} \resizebox{0.30\textwidth}{!}{\includegraphics{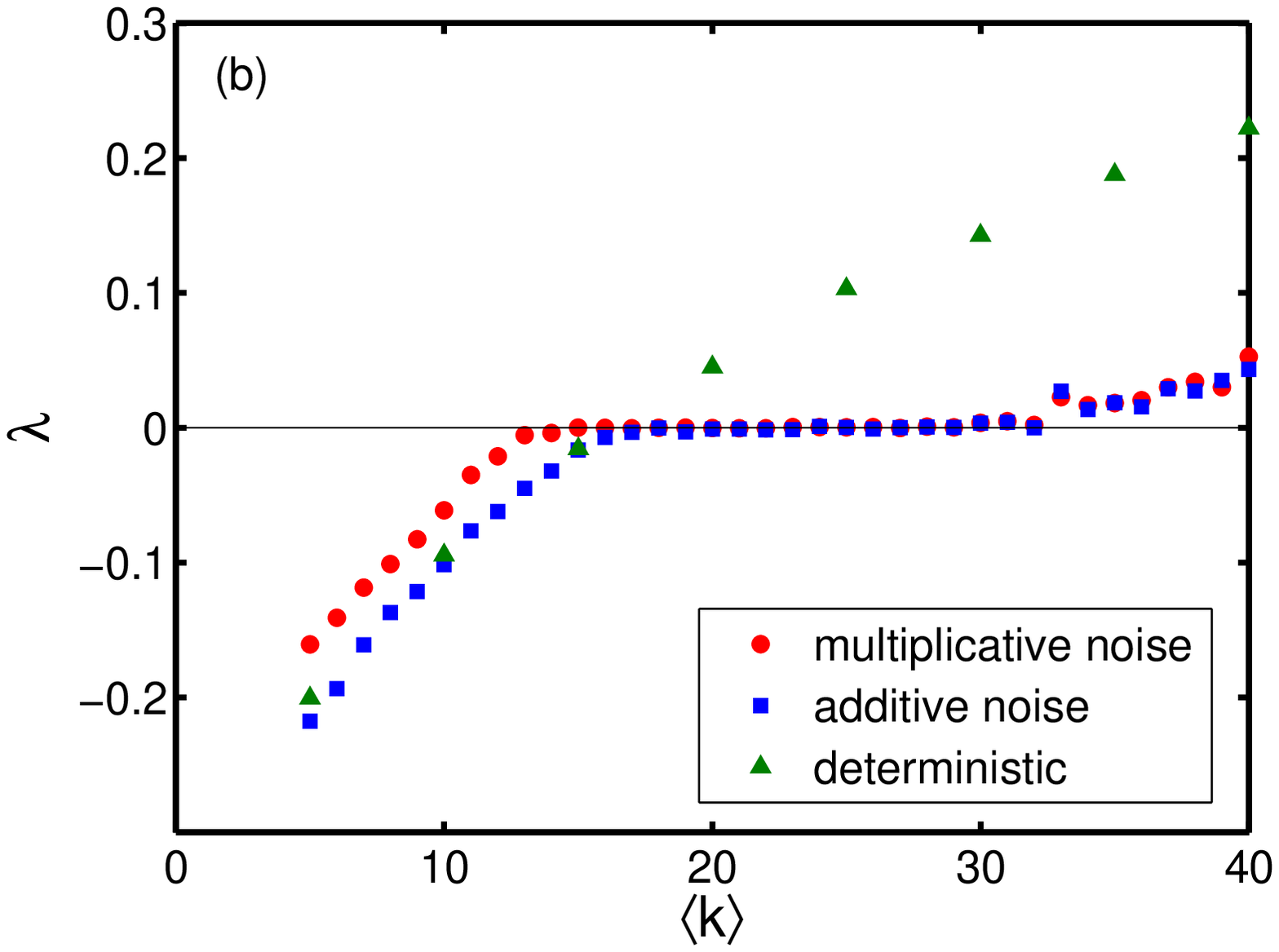}} \\
\hspace{0.5cm} \resizebox{0.30\textwidth}{!}{\includegraphics{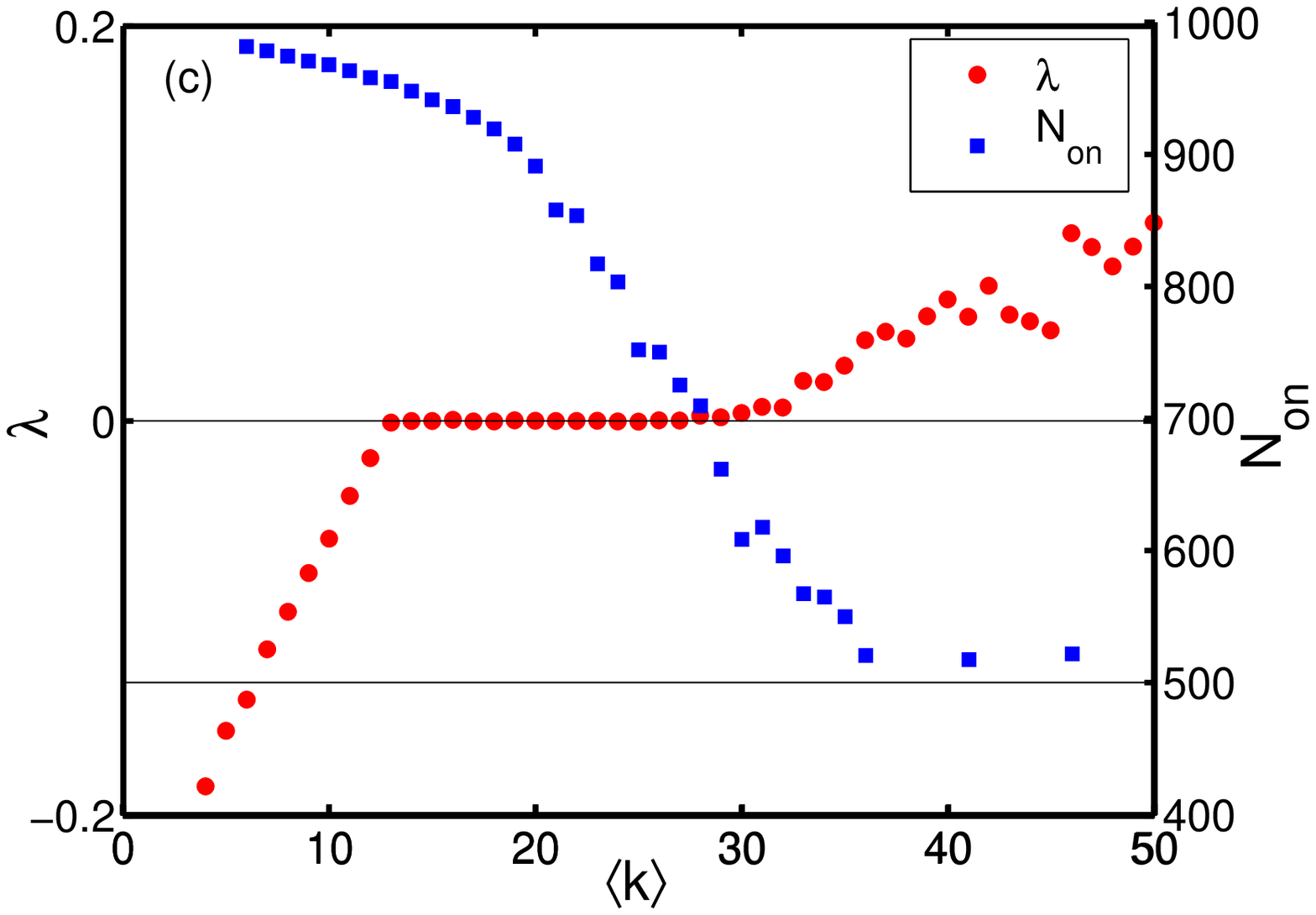}} \\
\hspace{-0cm} \resizebox{0.32\textwidth}{!}{\includegraphics{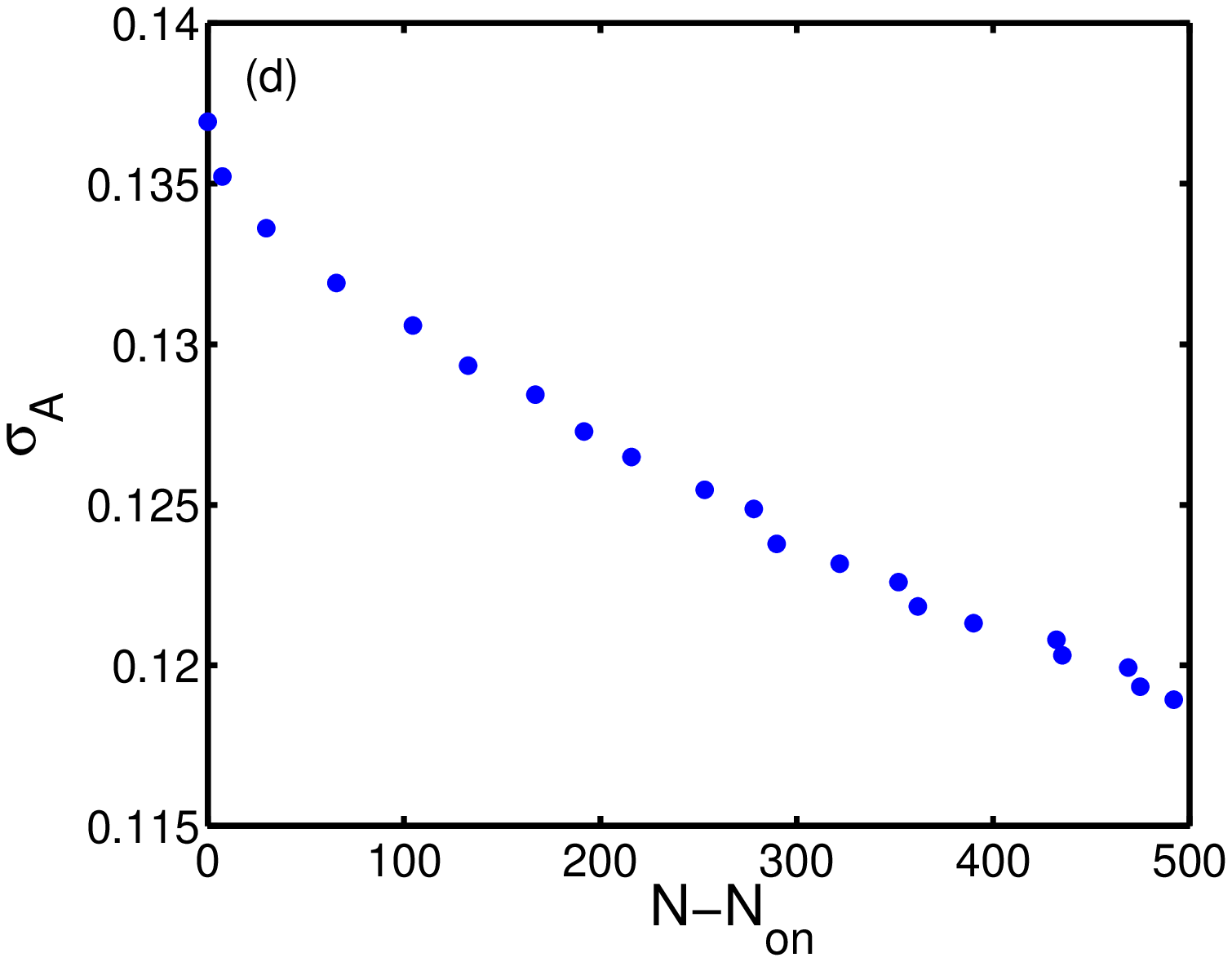}} 
\end{center}
\caption{
Lyapunov exponents for same parameters as in previous figure, for
(a)   different network types and sizes $N=200,500,1000$, 
(b) noise effects for multiplicative noise ($\bar \sigma = 0.1$ and $\sigma=0$), 
additive noise ($\bar \sigma = 0$ and $\sigma=0.1$), compared to the deterministic process ($\bar \sigma = \sigma=0$).
(c) $\lambda$ compared to the  number of inactive nodes as a function of  connectivity. ER, $N=1000$, $A_{ii}=-0.4$, 
and $\sigma=\bar \sigma=0.1$. 
(d)  Clearly $\sigma_A$ is not a constant, and declines with inactive nodes, 
as expected. ER, $N=1000$, $A_{ii}=-0.4$, and $\sigma=\bar \sigma=0$. 
}
\label{lyapfig}
\end{figure}

The stability of the system for different network topologies, sizes and  various  noise components 
is shown in Fig.  \ref{lyapfig}.
Figure  \ref{lyapfig}a  indicates that both, network size and degree distribution are slightly influencing 
the width of the plateau.
While in the $\langle k\rangle \rightarrow N$ region there is no significant difference in system stability, 
the low connectivity region shows  a size effect on the $\lambda=0$ plateau. 
The effect of network topology is relatively small, the curve pertaining to SF always being 
slightly below the ER networks, see Fig. \ref{lyapfig}a. 
While the width of the plateau is always wider for the random distribution of links, 
in the $\langle k\rangle \rightarrow 0$ region, the system is more stable (smaller $\lambda$) for SF networks. 
For higher connectivity regions ($\langle k\rangle \gtrsim 30$) the difference between random and scale-free networks
becomes indistinguishable due to numerical inaccuracy. 
Figure  \ref{lyapfig}b  shows the influence of pure multiplicative ($\bar \sigma >0, \sigma = 0$) and pure additive noise 
($\bar \sigma =0, \sigma > 0$) on the 
plateau width, compared to the deterministic process, $\bar \sigma =\sigma = 0$. 
With multiplicative noise the plateau becomes significantly wider, 
while additive noise hardly shows any effect when compared to the deterministic process.
Plateau widths are collected in Tab. 1.
\begin{table}
\begin{tabular}{c| c c c c c c c c c}
$A_{ii}$ $\backslash$ $\bar{\sigma}$ & 0 & 0.001 & 0.005 & 0.01 & 0.05 & 0.1 & 0.2 & 0.5 & 1\\
\hline
$-0.2$ & 4 & 3 & 4 & 5 & 5 & 5 & 4 & 2 & 0\\
$-0.4$ & 15 & 16 & 16 & 17 & 18 & 19 & 13 & 12 & 0\\
$-0.6$ & 15 & 14 & 15 & 16 & 20 & 23 & 24 & 18 & 0\\
\hline
$A_{ii}$ $\backslash$ $\sigma$ & 0 & 0.001 & 0.005 & 0.01 & 0.05 & 0.1 & 0.2 & 0.5 & 1\\
\hline
$-0.2$ & 5 & 6 & 5 & 4 & 5 & 4 & 4 & 4 & 4\\
$-0.4$ & 20 & 20 & 19 & 18 & 18 & 19 & 19 & 18 & 19\\
$-0.6$ & 23 & 22 & 23 & 22 & 21 & 23 & 21 & 22 & 21\\
\end{tabular}
\caption{Zero-$\lambda$ plateau widths $\Delta$ for an ER network, with $N=500$. The width $\Delta$ is defined as the
region of connectivity where $|\lambda|<0.005$. For the situation of
variable multiplicative noise the additive noise was fixed to $\sigma = 0.1$, for the variable additive noise,
$\bar{\sigma} = 0.1$. Cases for different $A_{ii}$ are shown.}
\end{table}

To understand the formation of the plateau ($\lambda =0$), one could naively expect  Eq. (\ref{maxlyap}) to hold 
at the plateau with the modification that $\langle k\rangle$ is replaced by  $\langle k_{\rm eff }\rangle$ from 
Eq. (\ref{maxlyap3}). This then would imply 
\begin{equation}
   N_{\rm on }=\frac{N A_{ii}^2 }{\sigma_A ^2 \langle k\rangle } \quad .
\label{expe}
\end{equation}
In Fig. \ref{lyapfig}c we show that the tail of $N_{\rm on } \sim -\langle k\rangle $ and does {\em not}  follow the naive 
expectation from Eq. (\ref{expe}), $\sim \langle k\rangle ^{-1}$. 
The key to understanding the formation of the plateau lies in the fact
that $\sigma_A$ is not a constant. When  $\lambda \to 0$, and nodes start becoming inactive, 
this amounts to effectively reducing the interaction matrix. For each product $i$ hitting zero, 
this means that row and column of matrix $A$  can be dropped. 
(In biological terms, once the concentration of some product  type reaches zero, 
these molecules stop playing a  role in the regulation of other products,  
the effective regulation network  consisting just of active 
nodes ($x_{i}>0$),  gets smaller.)
The key observation is that this does not leave the variance of the matrix elements $A_{ij}$ unchanged, 
but systematically reduces 
the variance the more nodes become inactive. This can be seen as a selection mechanism in 
which the most active reactions (largest reaction rates in $A$) will hit the boundary first, 
and will be the first ones to be  removed on average. This mechanism 
drives the system to a self-organized critical state at $\lambda=0$. 
We show the situation in Fig. \ref{lyapfig}d;  $\sigma _A$ declines with the number  of inactive nodes 
$N-N_{\rm on }$, as expected.
With further increasing $\langle k\rangle$ the number of inactive  products saturates  at 
half  the network size and  Eq. (\ref{maxlyap2}) holds.

\section{Discussion}

We have studied the stability of a simple stochastic model of catalytic reaction equations for 
cellular products such as mRNA molecules or proteins.
The system is driven by intrinsic molecular noise (multiplicative) and external (additive) noise. We 
show that the model captures some essential experimental features, such as the fat tail distribution
of concentration changes. Imposing an intuitively natural constraint on the system, (non-negativity of concentrations) 
we observe the forming of a plateau of vanishing Lyapunov exponents.  
The dynamical stability of  concentrations in catalytic regulatory networks, defined with 
Eq. (\ref{model}) has three extended phases in parameter space (here  connectivity). 
In the first phase the system is asymptotically stable, $\lambda$ is negative.  After being 
exposed to a random perturbation in this phase the system always relaxes to its steady state $x^0$. 
The main finding of this work is the existence of a  second phase, 
where $\lambda \sim 0$ extends over extended regions. This is in marked contrast to the dynamics of many other non-linear 
systems, which have $\lambda=0$ only at a set of points.
The emergence of this phase can be fully explained within the model. At higher connectivities some products 
will start to reach the boundary. Those products with the largest reaction rates will -- on average --  hit the 
boundary first. These are then removed from the system. This means that the variance of the effective 
reaction rates will get driven downward as a function of connectivity. The variance  of rates and the 
number of active nodes  balance each other to exactly arrive at the critical point $\lambda =0$. This is a 
self-organized critical effect. A physical  analogy is the evaporation at boiling temperature, where 
molecules of higher-than-average energy leave the liquid first, keeping the (critical) temperature at the boiling point. 
The third phase is defined by $\lambda > 0$ where system is dynamically unstable and  concentration levels are diverging. 
We studied the dependence of the $\lambda=0$ plateau on two network topologies, ER and SF, where a remarkably small  
effect was found. We found that with multiplicative noise the size of the plateau can be 
varied while additive noise showed relatively little effect. 
For strong enough noise levels of either type the plateau breaks down. 

In \cite{bertschinger} it was noted that neural networks can perform most 
complex computations if the dynamics of random threshold gate networks is {\emph at} the critical boundary 
between the ordered and chaotic regime. If we interpret  gene-regulatory networks as computing devices 
performing hundreds of optimization problems simultaneously, it is plausible that evolution would have selected 
among the most efficient variations -- working at the edge of chaos.

This project was supported by WWTF Life Science Grant LS 139, and by the Austrian Science Fund FWF, 
project P19132.

%\begin{thebibliography}{}\label{sec:TeXbooks}

\bibliographystyle{apsrev}

\end{document}